\documentclass[conference,letterpaper]{IEEEtran}

\usepackage[utf8]{inputenc} 
\usepackage[T1]{fontenc}
\usepackage{url}
\usepackage{ifthen}
\usepackage{cite}
\usepackage[cmex10]{amsmath} % Use the [cmex10] option to ensure complicance
\usepackage{textcomp,enumerate, url, cancel}                    			% Document and Style 
\usepackage[margin=0.8in]{geometry}
\usepackage{color}
\usepackage{amssymb,amsthm,bbm,bm,nicefrac}   				% Mathematical symbols
\usepackage{graphicx,epsfig}                                                               		% Graphics
\usepackage{multirow}
\newcommand{\ket}[1]{\left|#1\right\rangle}

\newcommand{\braket}[2]{\left\langle #1\lvert#2\right\rangle}
\newcommand{\abs}[1]{\left\lvert #1\right\rvert}

\newcommand{\dketbra}[1]{\left|#1\right\rangle\left\langle#1\right|}

\newcommand{\norm}[1]{||#1||}
\newcommand{\E}[2]{\mathbb{E}_{#1}\left[#2\right]}
\newcommand{\Pt}{\mathbf{P}_{t}}
\newcommand{\Rt}{\mathbf{R}_{t}}

\newcommand{\be}{\begin{equation}} 							
\newcommand{\ee}{\end{equation}}
\newcommand{\ba}{\begin{align}}
\newcommand{\ea}{\end{align}}
\newcommand{\bematrix}{\left(\begin{matrix}}
\newcommand{\ematrix}{\end{matrix}\right)}

\theoremstyle{definition}

\theoremstyle{theorem}

\theoremstyle{lemma}

\theoremstyle{proposition}

\theoremstyle{corollary}

\theoremstyle{observation}

\theoremstyle{remark}

\newcommand{\tr}[1]{\mathrm{tr}\left[#1\right]}

\def\cA{\mathcal A}

\def\cR{\mathcal R}
\def\cS{\mathcal S}

\begin{document}
\title{Reinforcement-learning calibration of coherent-state receivers on variable-loss optical channels} 

% %%% Single author, or several authors with same affiliation:
% \author{%
%   \IEEEauthorblockN{Stefan M.~Moser}
%   \IEEEauthorblockA{ETH Zürich\\
%                     ISI (D-ITET)\\
%                     CH-8092 Zürich, Switzerland\\
%                     Email: moser@isi.ee.ethz.ch}
% }

%%% Several authors with up to three affiliations:
\author{%
  \IEEEauthorblockN{Matias Bilkis}
\IEEEauthorblockA{\textit{Departament de F\'{\i}sica: Grup d'Informaci\'{o} Qu\`{a}ntica} \\
\textit{Universitat Aut\`onoma de Barcelona}\\
Bellaterra (Barcelona),
Spain \\
matias.bilkis@uab.cat}%
  \IEEEauthorblockN{Matteo Rosati}
\IEEEauthorblockA{\textit{Departament de F\'{\i}sica: Grup d'Informaci\'{o} Qu\`{a}ntica} \\
\textit{Universitat Aut\`onoma de Barcelona}\\
Bellaterra (Barcelona),
Spain \\
matteo.rosati@uab.cat}%
  \IEEEauthorblockN{John Calsamiglia}
\IEEEauthorblockA{\textit{Departament de F\'{\i}sica: Grup d'Informaci\'{o} Qu\`{a}ntica} \\
\textit{Universitat Aut\`onoma de Barcelona}\\
Bellaterra (Barcelona),
Spain \\
john.calsamiglia@uab.cat}

}

%%% Many authors with many affiliations:
% \author{%
%   \IEEEauthorblockN{Albus Dumbledore\IEEEauthorrefmark{1},
%                     Olympe Maxime\IEEEauthorrefmark{2},
%                     Stefan M.~Moser\IEEEauthorrefmark{3}\IEEEauthorrefmark{4},
%                     and Harry Potter\IEEEauthorrefmark{1}}
%   \IEEEauthorblockA{\IEEEauthorrefmark{1}%
%                     Hogwarts School of Witchcraft and Wizardry,
%                     1714 Hogsmeade, Scotland,
%                     \{dumbledore, potter\}@hogwarts.edu}
%   \IEEEauthorblockA{\IEEEauthorrefmark{2}%
%                     Beauxbatons Academy of Magic,
%                     1290 Pyrénées, France,
%                     maxime@beauxbatons.edu}
%   \IEEEauthorblockA{\IEEEauthorrefmark{3}%
%                     ETH Zürich, ISI (D-ITET), ETH Zentrum, 
%                     CH-8092 Zürich, Switzerland,
%                     moser@isi.ee.ethz.ch}
%   \IEEEauthorblockA{\IEEEauthorrefmark{4}%
%                     National Chiao Tung University (NCTU), 
%                     Hsinchu, Taiwan,
%                     moser@isi.ee.ethz.ch}
% }

\maketitle

%%%%%%
%% Abstract: 
%% If your paper is eligible for the student paper award, please add
%% the comment "THIS PAPER IS ELIGIBLE FOR THE STUDENT PAPER
%% AWARD." as a first line in the abstract. 
%% For the final version of the accepted paper, please do not forget
%% to remove this comment!
%%
\begin{abstract}
We study the problem of calibrating a quantum receiver for optical coherent states when transmitted on a quantum optical channel with variable transmissivity, a common model for long-distance optical-fiber and free/deep-space optical communication~\cite{Dequal2020,Andrews2005,Usenko2012a,Pirandola2021,Pirandola2021a,Vasylyev2011,Vasylyev2017}. 
  
  We optimize the error probability of legacy adaptive receivers, such as Kennedy's and Dolinar's~\cite{Kennedy1973a,Dolinar1973}, on average with respect to the channel transmissivity distribution. 
We then compare our results with the ultimate error probability attainable by a general quantum device, computing the Helstrom bound for mixtures of coherent-state hypotheses, for the first time to our knowledge, and with homodyne measurements. 

With these tools, we first analyze the simplest case of two different transmissivity values; we find that the strategies adopted by adaptive receivers exhibit strikingly new features as the difference between the two transmissivities increases. 
%Then we apply our methods to the study of a realistic model based on atmospheric turbulence~\cite{Vasylyev2011}.

  Finally, we employ a recently introduced library of shallow reinforcement learning methods~\cite{Bilkis2020}, demonstrating that an intelligent agent can learn the optimal receiver setup from scratch by training on repeated communication episodes on the channel with variable transmissivity and receiving rewards if the coherent-state message is correctly identified. 
\end{abstract}

%% The paper must be self-contained. However, if you are referring to
%% a full version for checking certain proofs, please provide the
%% publically accessible location below.  If the paper is completely
%% self-contained, you can remove the following line from your
%% submission.

\section{Introduction}
The study of information transmission using quantum states of light has laid the foundations of quantum information technologies, following the seminal works of Holevo, Helstrom and Bennet~\cite{holevoBOOK,helstromBOOK,Bennet1984}. Nowadays, the use of quantum devices to encode and decode information promises to boost long-distance communication rates~\cite{Banaszek2020,Rosati16b,Rosati16c} and provide unconditional security~\cite{Pirandola2019}. 

The simplest classical-information-transmission protocol on a quantum channel can be described in three steps: \begin{enumerate} 
\item classical information is encoded into a coherent state of the electromagnetic field, e.g., for a binary phase-shift keying (BPSK) modulation via a map \begin{equation}x\in\{0,1\}\mapsto\ket{\alpha_x},\end{equation} where the latter is a coherent state of amplitude $\alpha_x:=(-1)^x a\in\mathbb{R}$,  \begin{align}
&\ket\alpha:=\sum_{n=0}^\infty\frac{1}{\sqrt{n!}}e^{-\frac{\abs{\alpha}^2}{2}}\alpha^{n}\ket n,
\end{align}
with Poissonian photon-number statistics $p(n|\alpha)=\frac{1}{n!}e^{-\abs{\alpha}^2}\abs{\alpha}^{2n}$, and $\{\ket n\}_{n=0}^{\infty}$ is the Fock basis of the infinite-dimensional Hilbert space of one mode of the electromagnetic field.

\item the quantum state is sent through a lossy channel, which models the attenuation effect typical of long-distance optical-fiber, atmospheric and deep-space links, mapping \begin{equation}
{\cal L}_\eta:\ket \alpha\mapsto \ket{\sqrt\eta \alpha},    
\end{equation} 
where $\eta\in[0,1]$ is the transmissivity or attenuation coefficient of the lossy channel;
\item the received signal is measured by a generic quantum device, a positive operator-valued measurement (POVM), described by a set of operators ${\cal M}:=\{M_j\}_{j=0,1}$ such that $M_j\geq0$ and $\sum_j M_j=\mathbf{1}$, which outputs the estimate $j$, when measuring a state $\ket\alpha$, with probability $p(j|\alpha)=\tr{M_j\dketbra\alpha}$.
\end{enumerate}
The average success probability of this communication strategy, assuming that the signals are modulated with probability $q_x$, is
\begin{equation}
    P_{\rm s}({\cal M};\eta):=\sum_{x=0,1}q_x p(x|\sqrt\eta\alpha_x).
\end{equation}
This quantity, optimized over all quantum POVMs, leads to the celebrated Helstrom bound~\cite{helstromBOOK,Osaki1996}. For a BPSK coherent-state constellation, the latter bound can be approached in practice by a measurement device comprising adaptive linear optics and a threshold photodetector, as predicted by the seminal works of Kennedy and Dolinar~\cite{Kennedy1973a,Dolinar1973} (see~\cite{Cook2007} for a modern proof-of-principle demonstration of Dolinar's receiver), beating the performance of standard homo-/hetero-dyne techniques~\cite{RevGauss}.

However, practical communication links are often affected by noise-parameter variations, which can be difficult to estimate and counteract in real-time and deviate from simple theoretical models. In particular, in the case of a lossy channel ${\cal L}_\eta$, the transmissivity $\eta$ can be altered over time, a phenomenon sometimes called fading, which is caused by a plethora of different effects that alter the optical signal during its transmission through the atmosphere to/from a satellite~\cite{Dequal2020,Andrews2005,Usenko2012a,Pirandola2021,Pirandola2021a,Vasylyev2011,Vasylyev2017}. In this setting, the performance of all the strategies mentioned above can be expected to degrade, but a quantitative analysis and comparison of their behaviour has never been undertaken so far (see however~\cite{Cacioppo2021} for a preliminary transmission-rate analysis via the compound channel framework). 

In this work, we provide a first systematic study of one-shot coherent-state discrimination strategies in the presence of variable-loss channels: restricting to the case of two lossy channels $\{{\cal L}_{\eta_0}, {\cal L}_{\eta_1}\}$, happening with probabilities $\{\pi_0=\pi$, $\pi_1=1-\pi\}$, we take as a figure of merit the average success probability over the possible channel realizations,
\begin{equation}\label{eq:meanps}
    \bar P_{\rm s}({\cal M}):=\sum_{i=0,1}\pi_i  P_{\rm s}({\cal M};\eta_i).
\end{equation}
We compute the Helstrom bound for this average discrimination problem and evaluate the performance of a homodyne receiver. Then, we move on to evaluate the performance of an adaptive receiver (AR) with $L$ discrete adaptation steps, which includes the Kennedy receiver for $L=1$ and approximates the Dolinar receiver ($L\rightarrow\infty$) for large $L$: we compare the optimal success probabilities attainable by one and two-layer receivers with that of the homodyne receiver and the Helstrom bound across different energy regimes.

Finally, employing a recently developed library of shallow reinforcement learning methods~\cite{Bilkis2020}, we study the receiver's practical adaptation capabilities by training an intelligent agent on repeated discrimination experiments, where the channel transmissivity can change at each experimen. The agent is able to choose a receiver setup and guessing rule, and it is rewarded for each correct identification. We hence demonstrate that the agent is able to calibrate the receiver, i.e., discover a near-optimal setup, by trial-and-error.  
This result suggests the use of artificial intelligence for realizing a self-calibrating AR able to adapt to variable channel conditions in an experimental setting, without the need to estimate the variable channel parameters. 

The actual timescales of the learning process are dependent on the electronics employed in the implementation and are thus left open for further studies. We stress that the channel studied here is memoryless, hence the study of techniques which correlate multiple channel uses to estimate the time-local transmissivity are beyond the scope of the paper.

 \section{Discriminating binary coherent states on variable-loss channels}
\subsection{Ultimate performance: the Helstrom bound}
The Helstrom bound~\cite{helstromBOOK} determines the minimum error probability (or, equivalently, the maximum success probability) attainable in the discrimination of two quantum states with the most general quantum measurement. Its value is well-known for binary coherent states, and it is achieved by a projection on a superposition of the states~\cite{Osaki1996}. However, in our case each hypothesis state can take two different values, depending on the channel transmissivity. This effectively amounts to discriminate between the two mixed states 
\begin{equation}
\rho_{x}:=\sum_{i=0,1}\pi_{i}\dketbra{\sqrt{\eta_{i}}\alpha_{x}}, \;\;\;x\in\{0,1 \} .
\end{equation}
The average success probability of a generic measurement then reads
\begin{equation}
\bar P_{\rm s}({\cal M})=\sum_{x}q_{x}\tr{M_{x}\rho_{x}}
\end{equation}
and its optimal value is given by the Helstrom bound
\begin{equation}\label{eq:helstrom}
\bar P_{\rm s}^{\rm opt}=\frac{1+\norm{q_{0}\rho_{0}-q_{1}\rho_{1}}_{1}}{2},
\end{equation}
where $\norm{\cdot}_{1}={\rm Tr}|\cdot|$ is the operator trace-norm.

To compute the Helstrom bound in our case, we compute the trace-norm of the matrix $q_0\rho_0-q_1\rho_1$ in the four-dimensional subspace where it is supported, ${\cal S}:={\rm span}\{\ket{\psi_{x+2i}}:=\ket{\sqrt{\eta_{i}}\alpha_{x}}\}_{i,x=0,1}$. Since the trace-norm is basis-independent, one can readily find a four-dimensional representation of the involved states by computing the square-root of their Gram matrix, $B=\sqrt{G}$ with $G_{ij}=\braket{\psi_i}{\psi_j}$. By construction it holds that $B^\dagger B= G$, hence the column vectors of $B$ constitute a representation of the original states in a $4$-dimensional subspace of the Hilbert space. By expressing all the operators via this representation, we are then able to compute \eqref{eq:helstrom} numerically.

 \subsection{The adaptive receiver}
The AR considered in this paper is a generalization of Kennedy's and Dolinar's receivers~\cite{Kennedy1973a,Dolinar1973}, where the incoming signal is sequentially split into $L$ distinct fractions via repeated interaction with a beam-splitter of transmissivity $\theta_{\ell}$, for $\ell=1,\cdots,L$. Each extracted fraction is manipulated via an optical displacement of amplitude $\beta$, mapping
\begin{equation}
D(\beta):\ket\alpha\mapsto\ket{\alpha+\beta},
\end{equation}
and then measured via a threshold photodetector, with a binary outcome $j=0,1$, corresponding to detecting respectively no photons, one or more photons. The resulting single-layer POVM is ${\cal M}_{\rm ken}:=\{\dketbra\beta,\mathbf1-\dketbra\beta\}$. The outcome probability at layer $\ell$ then is
\begin{equation}
p(0|\alpha^{(\ell)})=e^{-\abs{\alpha^{(\ell)}-\beta}^{2}},~p(1|\alpha^{(\ell)})=1-p(0|\alpha^{(\ell)}),
\end{equation}
with $\alpha^{(\ell)}:=\sqrt{\theta_{1}\cdots\theta_{\ell-1}(1-\theta_{\ell})}\alpha$ the amplitude extracted at the $\ell^{\text{\underline{th}}}$ layer for detection, for a coherent-state input. In our case, the outcome probability will be an average with respect to the signals obtained with different transmissivities: 
\begin{equation}
p(j|\rho^{(\ell)}_{x}):=\sum_{i=0,1}\pi_{i}p(j|\sqrt{\eta_{i}}\alpha^{(\ell)}_{x}).
\end{equation}

The receiver is adaptive because it determines the displacement at the next layer, $\beta^{(\ell+1)}$, based on the history of outcomes $j^{(\ell)}$ and displacements $\beta^{(\ell)}$ at each previous layer. Moreover, after the last detection, the optimal guess is given according to the maximum-likelihood criterion, based on the posterior probabilities for the hypotheses. The overall success probability can be written as
\begin{equation}
\bar P_{\rm s}^{\rm ar}:=\sum_{\ell=1}^{L}\max_{x=0,1}q_{x}p(j^{(L)}\cdots j^{(1)}|\rho_{x}).
\end{equation}
The Kennedy receiver is obtained for the single-transmissivity case by setting $L=1$, $\theta_{1}=0$ and $\beta^{(1)}=\pm\alpha_{x}$, while the Dolinar receiver by taking infinitesimal steps $\theta_{\ell}=1/L$, sending $L\rightarrow\infty$ and a suitable choice of the adaptive strategy $\beta^{(\ell)}$~\cite{Dolinar1973}. Strikingly, the latter attains the Helstrom bound for BPSK discrimination and was demonstrated in~\cite{Cook2007}. General AR's can be obtained for finite steps~\cite{Bilkis2020}, multiple hypotheses~\cite{Muller2015,Rosati2017a,Rosati2017b} and non-linear operations in the branch~\cite{Rosati16a}, but are not guaranteed in general to attain the corresponding minimum-error probability~\cite{Nakahira2018b}. 

In our case, there is also no guarantee that the AR will be able to attain the Helstrom bound. In the following sections we investigate the performance of this receiver family for the discrimination of coherent-state mixtures arising from the practically motivated condition of variable channel loss. 
 
 \subsection{Homodyne receiver}
 Another common coherent-state discrimination strategy employs homodyne detection, corresponding to a measurement of the $\hat q:=\frac{1}{2}(\hat{a}+\hat{a}^\dagger)$ quadrature of the field, i.e., ${\cal M}_{hom}:=\{\dketbra q\}_{q\in\mathbb R}$. It is known to be suboptimal with respect to an AR, but it is much simpler to implement and performs reasonably well for BPSK~\cite{RevGauss,Takeoka2008}. In our setting, it might provide a further advantage since it is not immediately affected by the variable attenuation, given that it can distinguish more than two amplitude values. Using the fact that $p(q|\alpha)=\sqrt{\frac{2}{\pi}}e^{2(q-\alpha)^2}$ and that the optimal decision rule is $x=0$ ($x=1$) for $q\geq 0$ ($q<0$) one readily finds that homodyne success probability reads
 \begin{equation}\label{eq:homodyne}
% \bar P_{\rm e}^{\rm hom}=\frac{1}{2}\sum_{i=0,1}\pi_{i}{\rm Erfc}\left[{\sqrt{2\eta_{i}} a }\right]\right).
% \end{equation}
 \bar P_{\rm s}^{\rm hom}=\frac{1}{2} \big( 1 + \sum_{i=0,1}\pi_{i}{\rm Erf}\left[{\sqrt{2\eta_{i}} a }\right]\big).
 \end{equation}
 
 \section{Comparing different receivers}
\begin{figure}[t!]
    \centering
    \includegraphics[width=.5\textwidth]{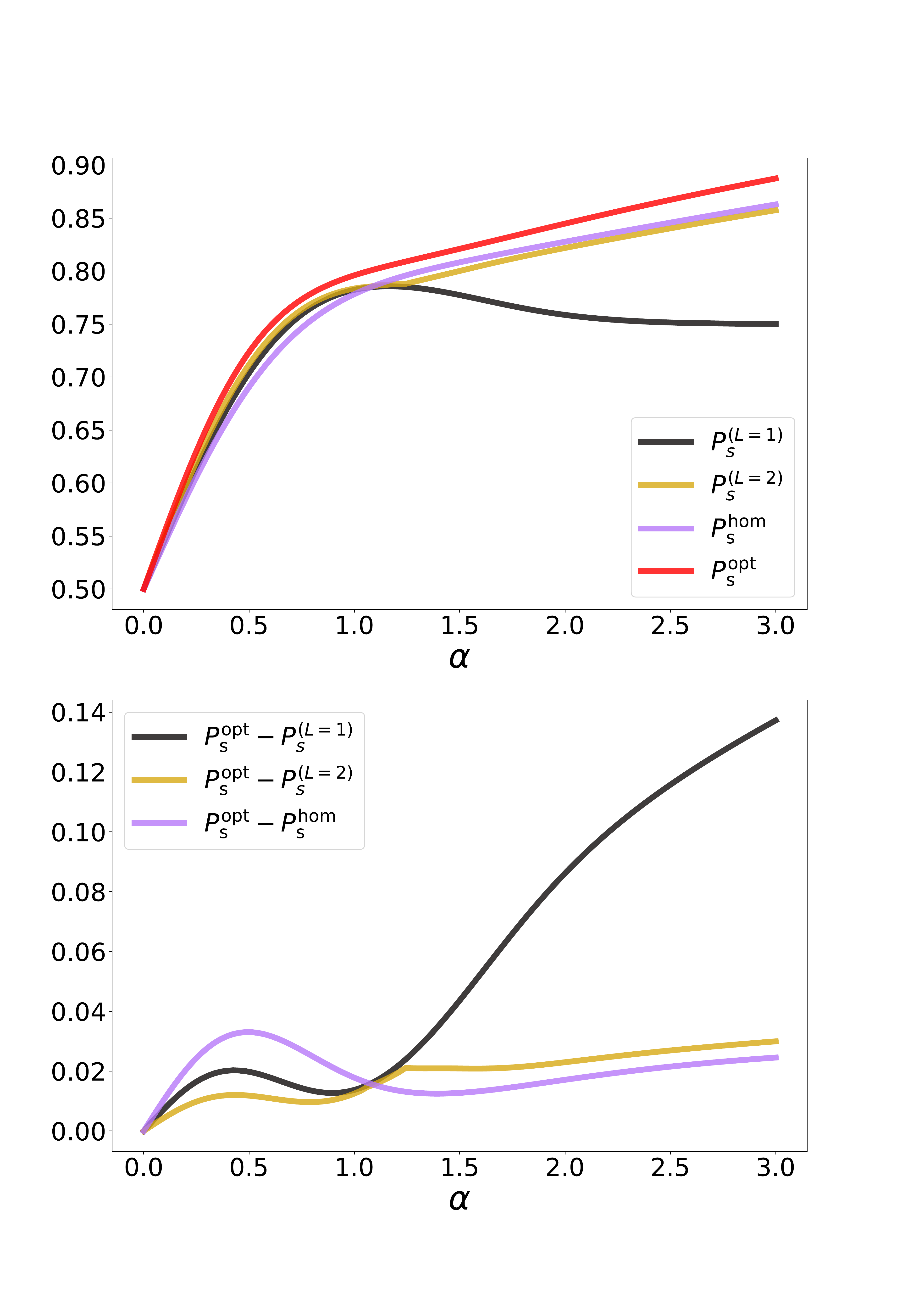}
    \caption{Top: We compare the success probabilities attainable by one $(L=1)$ and two ($L=2$) adaptive receivers with homodyne measurement and Helstrom bound, for different values of signal intensity. Bottom: difference between the success probabilities of the receivers under considerations and the Helstrom bound are shown in order to ease the visualizaton of different features. In this case, a lossy channel ($\eta=10^{-2}$) acts on the transmitted state with probability of one half, whereas the signal is transmitted through a noise-less channel otherwise. }
    \label{fig:benchmark}
\end{figure}

 We now compare the success probabilities attained by the different receivers introduced above. In particular, we consider the parameter regime where the two transmissivities are relatively distant from each other, which approximates the physical cases where the channel transmissivity changes abruptly due to weather conditions~\cite{Vasylyev2017}. In Fig.~\ref{fig:benchmark} we compare the behaviour of Kennedy, 2-layer adaptive and homodyne receivers across different values of input signal amplitude, together with the Helstrom bound. The success probability of the homodyne receiver is given by Eq.~\eqref{eq:homodyne}, whereas displacements in the adaptive receivers have been numerically optimized using simulated annealing. This optimization proves demanding for higher number of adaptive measurements and whether this kind of receivers attain the Helstrom bound for $L\rightarrow \infty$, when transmitting signals under the presence of two lossy-channels, still remains an open question.
 
 We observe that, while the homodyne receiver maintains a satisfying performance in the whole amplitude range, the behaviour of the Kennedy receiver exhibits a clear transition and its performance degrades at sufficiently large amplitudes. Fortuantely, the addition of a second adaptive detection layer appears to be able to correct this trend. Hence the two-layer receiver can get close to the performance of the homodyne receiver in the large-amplitude regime, while showing a clear advantage with respect to the latter in the low-amplitude regime. 
 By inspection of the optimal displacements chosen by the one- and two-layer receivers, we observe that the transition takes place when the two possible received signal amplitudes, $a_0:=\eta_0 |\alpha|$ and $a_1:=\eta_1 |\alpha|$, are so distant that a displacement nullying the largest of them, say $a_1$, as prescribed by the optimal Kennedy receiver, necessarily increases the probability that the other possible signal, $a_0$ clicks. This happens roughly when $a_1-a_0\simeq a_0$. Further analyses will be carried out to study the actions taken at later layers to correct this behaviour.

 Having compared the success probabilities attainable by the different receivers, we now turn to study how machine-learning agent can deal with this problem under a more realistic scenario, at which nor channel parameters neither outcomes probabilities need not to be a priori known, and an adaptive receiver needs to be optimized based solely on measurement outcomes. To this end, we tackle this problem as a Reinforcement Learning one.

 \section{Machine learning framework}

Reinforcement Learning is a branch of Machine Learning that can be understood under the framework of sequential decision-making processes, in which an agent and an environment interact during several episodes~\cite{Sutton2018}. %In this work, we use ideas from reinforcement learning to show that a model-agnostic agent can learn to calibrate an AR by using only real-time measurement outcomes. 

At each time-step $\ell=0,\cdots,L$ of each episode $t=1,\cdots,T$, the agent observes the environment in a \textit{state} $s_{\ell}^{(t)}\in\cS$ and chooses an \textit{action} $a_\ell^{(t)}\in\cA$; as a consequence, the agent enjoys a reward $r_{\ell+1}^{(t)}\in\cR$ and observes a new state of the environment, $s_{\ell+1}^{(t)}\in\cS$; where $\cS$, $\cA$ and $\cR$ stand for the sets of states, actions and rewards the agent may experience.
 
Environment dynamics is completely determined by the \textit{transition function} $\tau(s',r|s,a)$, i.e., the conditional probability of ending up in a state $s'$ and conferring a reward $r$, given that the previous state was $s$ and the agent took an action $a$; the next future states accessible from $s$ are thus restricted to $\cS(s) = \{ s' : \tau(s'| s) \neq 0 \} \subseteq \cS$. 

While the agent does not have control of nor access to the transition function, it will influence the dynamics of the environment by choosing actions according to an interaction \textit{policy} $\pi(a|s)$, i.e., the conditional probability of performing an action $a$ when the observed environment's state is $s$. In case this observation provides a complete description of the environment state, e.g. $s_{\ell}$, the setting is known as a Markov decision process (MDP). 

The agent's objective is to acquire as much reward as possible during an episode. As a matter of fact this strongly depends on the agent's policy. By introducing the return, $G_{\ell}^{(t)}=\sum_{i=0}^{L-\ell}\gamma^i r_{i+\ell+1}^{(t)}$, it is straightforward to quantify how valuable it is to take action $a$ at a given state $s$ and following the policy $\pi$ afterwards; this is done via the so-called state-action value function $Q_\pi(s)=\E{\pi}{G_\ell | s_\ell=s, a_\ell = a}$, where the expectation is taken over all possible trajectories obtained by departing from state $s$ and performing action $a$.

By writing explicitly the expected value for the first future time-step, Bellman equation follows~\cite{Bellman2003}:
{\small\begin{eqnarray*}
Q_{\pi}(s,a)&=& \sum_{\substack{ s'\in\cS, r\in\cR\,\\ a\in\cA} } \tau(s',r|s,a)(r+\gamma \pi(a'|s')Q_{\pi}(s',a')).\nonumber
\end{eqnarray*}}
The decision problem can then be solved by finding an optimal policy $\pi^{*}$, whose associated state-action value function satisfies the optimal Bellman equation:
{\small\begin{eqnarray}\label{eq:qBellOp}
Q^{*}(s,a)&&:=Q_{\pi^*}(s,a)=\max_\pi Q_\pi(s,a)\\
&&=\sum_{s'\in\cS,r\in\cR}\tau(s',r|s,a)(r+\gamma \max_{a'\in\cA}Q^{*}(s',a')).\nonumber
\end{eqnarray}}

When accurately estimated via several agent-environment interactions, the state-action value function permits the agent to modify its policy towards the optimal one in a model-free way; in turn this is the spirit of the algorithm we next present.

\subsection{Q-learning}
Q-learning~\cite{Watkins1989} is an algorithm based on the observation that any Bellman operator, i.e. the operator describing the evolution of a state-action value function as in Eq.~(\ref{eq:qBellOp}) is contractive. Thus, in order to find $Q^{*}(s,a)$, Q-learning turns the optimal Bellman equation for $Q$, Eq.~\eqref{eq:qBellOp}, into an update rule for $\hat{Q}(s_{\ell},a_{\ell})$, i.e., the $Q$-function's estimate available to the agent at a given time-step $\ell$ of any episode $t=1,\cdots,T$. 

An interaction step is described by the tuple $s_{\ell}\rightarrow a_{\ell}\rightarrow r_{\ell+1}\rightarrow s_{\ell+1}$, and obtained by following an $\epsilon$-greedy interaction policy, e.g. that one which selects an action at random with probability $\epsilon$ or chooses the action maximizing current estimate of $\hat{Q}(s_\ell, a_\ell)$ otherwise. After an interaction step occurs, the update rule for the $Q$-estimate is
{\small\be\begin{aligned} \label{eq:QLUPDATERULE}
\hat{Q}(s_{\ell}, a_{\ell}) & \leftarrow (1-\tilde{\alpha})\hat{Q}(s_{\ell}, a_{\ell})\\
&+ \tilde{\alpha} \left(r_{\ell+1}  + \gamma \max_{a'\in\cA(s_{\ell+1})}\hat{Q}(s_{\ell+1}, a')\right),
\end{aligned}\ee}%
where $\tilde{\alpha}$ is the learning rate. Note that the update at each time-step $\ell$, requires only to enjoy the next immediate reward $r_{\ell+1}$ and to observe the next state $s_{\ell+1}$. Combined with the model-free prescription for the evolution of the Q-estimates, Q-learning constitutes a powerful tool for on-line learning of MDPs. 
 
 Motivated by the model-free and experimentally appealing features of the Q-learning algorithm, we now turn to apply it for the problem of calibrating a two-layer adaptive receiver in the regime where an advantage with respect to the homodyne one is present. Importantly, we will do so by assuming no knowledge on the lossy-channel parameter, signal intensity nor outcomes probabilities.

 \section{Learning a near-optimal receiver via Q-learning}\label{ssec:qLRes}
We now present the results obtained by a RL agent based on Q-Learning with $\epsilon$-greedy interaction policy. The setting is modelled as a Partially Observable Markov Decision process (POMDP), which can be reduced to an effective MDP by
defining an effective state that contains all the past history of observations and actions up to a given time-step, i.e., $h_{\ell}=(a_0,o_{1},\cdots,a_{\ell-1},o_\ell)$. Here, actions $a_{\ell}$ correspond to both choosing the displacement at each receiver layer $\ell$ and performing a guess for the incoming signal after the final observation is obtained; observations thus correspond to photodetection outcomes.

We evaluate the performance of the agent using two figures of merit as a function of the number of episodes elapsed so far, $t$: (i) the cumulative return per episode
\be
\mathbf{R}_{t}=\frac1t\sum_{i=1}^{t}G^{(i)}_{0}=\frac1t\sum_{i=1}^{t}r_{L+1}^{(i)},
\ee
where $r_{L+1}^{(i)}=\{1,0\}$ stands for the correctness of the guess made at episode $i$, and (ii) the success probability of the best actions according to the agent, at the current episode,
\be
\mathbf{P}_{t}=P_{s}(\alpha,\{a_{\ell}^{(t)*}\}),
\ee
where the best actions $\{ a_{\ell}^{(t)*} \}$ at episode $t$ are those which maximize current agent's $Q$-estimate.

%In contrast to the model-aware case, where the guessing rule is straightforwardly obtained from the Bellman equation at the last time-step, the model-agnostic optimization problem includes a non-trivial search for the optimal guessing rule, determined by the optimal $Q$-function.

Note that $\Rt$ can be assessed by the agent in real-time, and captures how well it acts as episodes occur: high values of this quantity not only prove that a good policy has been learnt, but also that it is being performed. %Algorithms aiming to optimize $\Rt$ follow a philosophy of learning-by-acting, that in turn allows the agent to guide the search over the state-action space. 
%This feature is captured by the evolution of $\Rt$ over different episodes: 
Thus, a good learner is that which gets as much reward as possible \textit{during} the learning process.
On the other hand, the second figure of merit, $\Pt$, is standard in Quantum State Discrimination, and in our context evaluates the best strategy discovered by the agent so far.

\begin{figure}[t!]
    \centering
    \includegraphics[width=.5\textwidth]{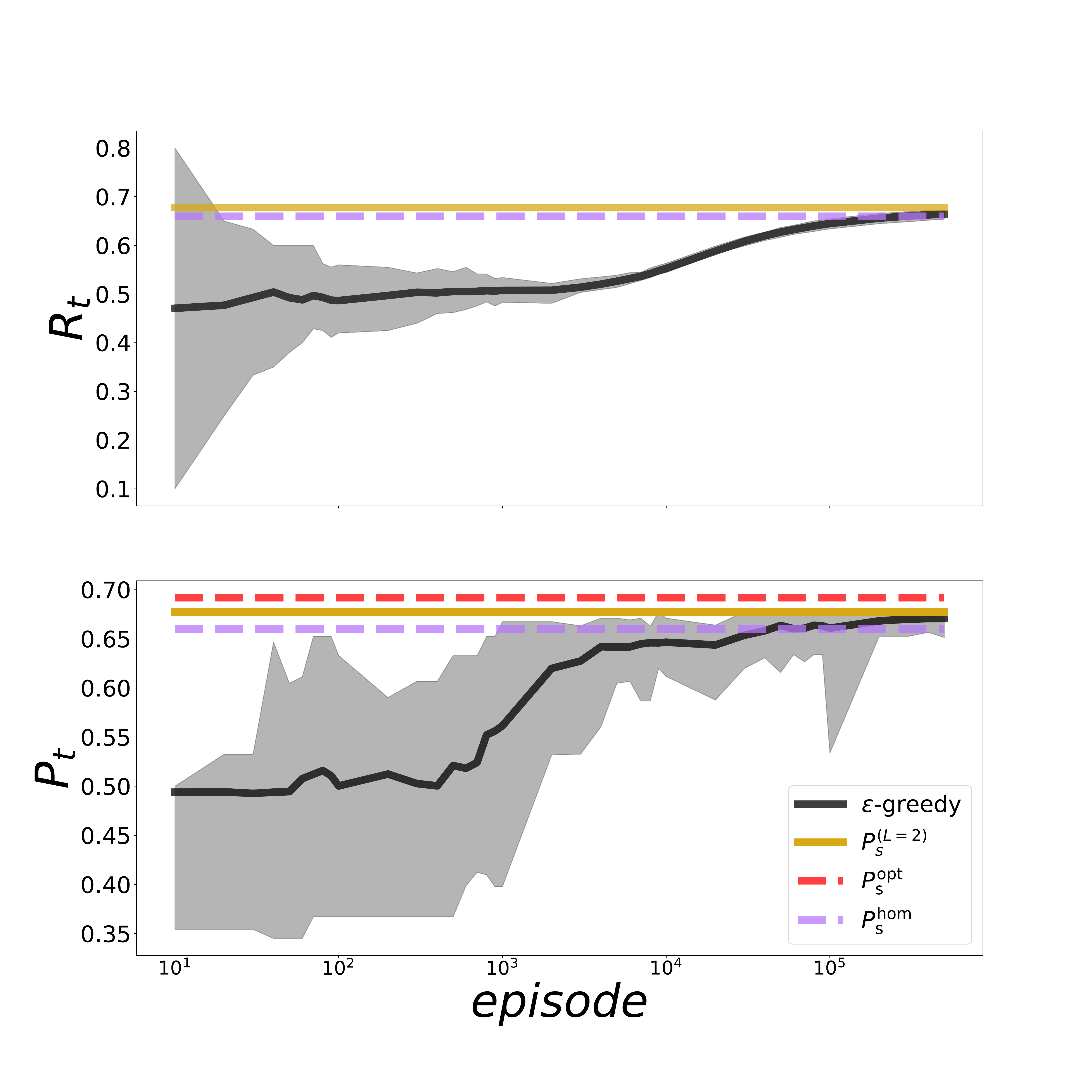}
    \caption{We show learning curve behaviour for $\epsilon$-greedy Q-learning, with an $\epsilon$ exponentially decaying as a function of episode number. The figures of merit are averaged over $A=24$ agents and corresponding uncertainty region are shown. }
    \label{fig:epgreedy}
\end{figure}

In Fig~\ref{fig:epgreedy} we show the learning curves for a model-agnostic RL agent, trained over $5 \; 10^{5}$ episodes. At each episode, a bit of information is encoded in a coherent-state (we fix $\alpha$=0.4) and transmitted; with probability of one half the signal gets attenuated by a lossy channel of parameter $\eta = 10^{-2}$ in the path between sender and receiver, or remains unchanged otherwise. The AR considered consists on 2 photodetections layers; a guess for the encoded information bit is done after the final measurement outcome is obtained, and a reward of 1 is given if the guess is correct, and zero reward otherwise. In this case, the agent is asked to decide which displacement to perform at each stage, and to finally guess for the bit value; a discretization among 10 displacements was taken between each photodetection layer, which constitues a non-trivial search among $~10^{3}$ configurations, each configuration leading to a binary reward. We observe that by the end of the training, the RL agent not only manages to discover a configuration that beats the homodyne receiver, but also to empirically surpass that bound in terms of $\Rt$. Moreover, we depict the evolution of the best strategy discovered by the agent, which is given by $\Pt$ and is straightforwardly related with the success probability $\bar{P}_{\rm s}(\mathcal{M})$, as defined in Eq.~\eqref{eq:meanps}.

This result opens up the door to investigate real-time calibration of coherent-state receivers transmitted under variable-loss channels, by means of machine learning methods. We stress that our simulations do not take into account actual delays present in a real-time situation, which go beyond the scope of the present preliminary investigation.

\section{Conclusions}
In this work, we have studied one-shot coherent-state discrimination strategies in the presence of variable-loss channels. We first computed the ultimate success probability attainable by a quantum receiver, i.e. the Helstrom bound, and compared it with that attained by homodyne receivers and Dolinar-like ones. To the best of our knowledge, this is the first time that a performance comparison among different classes of quantum receivers is given for this communication problem. Our results open up several questions in this regard; for example it is of great relevance (both technologically and theoretically) to understand in which parameter regimes Dolinar-like receivers can beat standard measurements (such as homodyne-like ones) and provide a quantum advantage even in the presence of noise parameter fluctuations, as seen in Fig.1. Furthermore, an important step to study more realistic scenarios will be to consider parameter fluctuations inside an interval of values, rather than two extreme points. This will be done by going beyond the two-transmissivities case, which is entirely possible with the instruments presented here by numerical means, introducing realistic transmissivity distributions~\cite{Vasylyev2011}. We expect that the behaviour of the system in this case will be dominated by the distribution's variance, determining an effective minimum and maximum transmissivity that can be analyzed as in this paper.
 Finally, the real-time performance of our techniques should be compared with that of refined techniques that make clever use of finite characteristic time for the parameter variations, as well as taking into account realistic delay times. 

%At variance with the decoding of a classical variable channel, which was tackled in~\cite{Farsad2017,Shlezinger2019} by the usage of neural networks, communication over quantum channels introduce an additional layer of complexity (and potential gain) due to the optimization of the quantum measurement device that extracts the classical signal. 

\section*{Acknowledgment}
MR acknowledges J. N\"{o}tzel and A. Cacioppo for suggesting the optimization of a Dolinar receiver in the framework of compound channels, see~\cite{Cacioppo2021}. This project has received funding from the European Union’s Horizon 2020 research and innovation programme under the Marie Skłodowska-Curie grant agreement No 845255. MB and JCC acknowledges support from the Spanish Agencia Estatal de Investigación, project PID2019-107609GB-I00, Generalitat de Catalunya CIRIT 2017-SGR-1127.

%%%%%%
%% To balance the columns at the last page of the paper use this
%% command:
%%
%\enlargethispage{-1.2cm} 
%%
%% If the balancing should occur in the middle of the references, use
%% the following trigger:
%%
%\IEEEtriggeratref{4}
%%
%% which triggers a \newpage (i.e., new column) just before the given
%% reference number. Note that you need to adapt this if you modify
%% the paper.  The "triggered" command can be changed if desired:
%%
%\IEEEtriggercmd{\enlargethispage{-20cm}}
%%
%%%%%%

%%%%%%
%% References:
%% We recommend the usage of BibTeX:
%%
%\bibliographystyle{IEEEtran}
%\bibliography{definitions,bibliofile}
%%
%% where we here have assume the existence of the files
%% definitions.bib and bibliofile.bib.
%% BibTeX documentation can be obtained at:
%% http://www.ctan.org/tex-archive/biblio/bibtex/contrib/doc/
%%%%%%

%% Or you use manual references (pay attention to consistency and the
%% formatting style!):
 \bibliographystyle{IEEEtran}
   \bibliography{/Users/matteorosati/Documents/Ricerca/Appunti/library}

\end{document}